\documentclass[aps,pra,twocolumn,floatfix,footinbib,groupedaddress,superscriptaddress]{revtex4-1}
\usepackage{amsmath}
\usepackage{amssymb,graphics,graphicx,times,bm,multirow,color}

\input{tcilatex}
\begin{document}

\title{Exciton-phonon information flow in the energy transfer process of
photosynthetic complexes}
\author{Patrick Rebentrost}
\email{rebentr@fas.harvard.edu}
\affiliation{Department of Chemistry and Chemical Biology, Harvard University, 12 Oxford
St., Cambridge, MA 02138}
\author{Al\'an Aspuru-Guzik}
\email{aspuru@chemistry.harvard.edu}
\affiliation{Department of Chemistry and Chemical Biology, Harvard University, 12 Oxford
St., Cambridge, MA 02138}
\pacs{PACS number}

\begin{abstract}
Non-Markovian and non-equilibrium phonon effects are believed to be key
ingredients in the energy transfer in photosynthetic complexes, especially
in complexes which exhibit a regime of intermediate exciton-phonon coupling.
In this work, we utilize a recently-developed measure for non-Markovianity
to elucidate the exciton-phonon dynamics in terms of the information flow
between electronic and vibrational degrees of freedom. We study the measure
in the hierarchical equation of motion approach which captures strong
system-bath coupling effects and non-equilibrium molecular reorganization.
We propose an additional trace-distance measure for the information flow that 
could be extended to other master equations.
We find that for a model dimer system and the Fenna-Matthews-Olson complex
that non-Markovianity is significant under physiological conditions.
\end{abstract}

\date{ \today }
\maketitle

\volumeyear{year} \volumenumber{number} \issuenumber{number} \eid{identifier}

\startpage{1}


The initial step in photosynthesis involves highly efficient excitonic
transport of the energy captured from photons to a reaction center \cite%
{Blankenship2002}. In most higher plants and other organisms such as green
sulphur and purple bacteria this process occurs in light-harvesting
complexes which consist of electronically coupled chlorophyll molecules
embedded in a solvent and a protein environment \cite{Cheng2009}. Several
recent experiments show that excitonic coherence can persist for several
hundreds of femtoseconds even at physiological temperature \cite%
{Engel2007,*Lee2007,*Panitchayangkoon2010,*Collini2010}. These experiments
suggest the hypothesis that quantum coherence may be biologically relevant
for photosynthesis. 

The challenging regime of intermediate coupling of the electronic to the
vibrational degrees of freedom motivates the study of sophisticated master
equations with non-Markovian effects. Several approaches have been taken,
ranging from a polaron transformation \cite{Jang2008}, quantum state
diffusion \cite{Roden2009}, generalized Bloch-Redfield \cite{Wu2010},
non-Markovian quantum jumps \cite{Piilo2008,Rebentrost2009b}, quantum path
integrals (QUAPI) \cite{Thorwart2009}, to density matrix renormalization
group (DMRG) \cite{Prior2010}. After photoexcitation, the nuclear
coordinates of a molecule will relax to new equilibrium positions, a
phenomenon that is borne out by the time-dependent Stokes shift, that is the
frequency shift between absorption and emission spectra. Redfield theory
assumes that the phonon bath is always in thermal equilibrium and thus
cannot capture this effect \cite{Ishizaki2009b}. Ishizaki and Fleming
employed the hierarchical equation of motion (HEOM) approach \cite%
{Ishizaki2009a,*Ishizaki2009}, which interpolates between the usual weak and
strong (singular) exciton-phonon coupling limits and takes into account
non-equilibrium molecular reorganization effects.

In this work, we study the non-Markovianity of the exciton transfer process
by means of numerical simulation. Quantum mechanical time evolution under
decoherence leads to transfer of information encoded in the electronic
excited state to the vibrational degrees of freedom. Non-Markovianity can be
characterized as the return of this information to the electronic degrees of
freedom. To quantify the exciton-phonon information flow, we employ the
state-of-the art master equation approach by Ishizaki and Fleming and a
newly-developed measure for non-Markovianity. The central question we will
answer is how much information is exchanged between excitonic and phononic
degrees of freedom in that process and, specifically, how much information
returns from the phonons to the exciton.

\textit{Measure for non-Markovianity}$-$ A recent area of research is the
study of measures to characterize non-Markovianity \cite%
{Breuer2009,Wolf2008,*Rajagopal2010,*Rivas2010}. We utilize the readily
applicable measure developed by Breuer and co-workers \cite{Breuer2009}. It
is based on the quantum state trace distance:%
\begin{equation}
D\left( \rho _{1}\left( t\right) ,\rho _{2}\left( t\right) \right) =\frac{1}{%
2}\text{Tr}\left\{ |\rho _{1}\left( t\right) -\rho _{2}\left( t\right)
|\right\} ,  \label{eqTraceDistance}
\end{equation}%
where $|A|=\sqrt{A^{\dagger }A}.$ This measure quantifies the
distinguishablility of two quantum states $\rho _{1}$ and $\rho _{2}$. For a
completely positive trace-preserving map $E$ the trace distance is
contractive, i.e. $D\left( E\left( \rho _{1}\right) ,E\left( \rho
_{2}\right) \right) \leq D\left( \rho _{1},\rho _{2}\right) $ \cite%
{Nielsen2000}. Time intervals in which $D$ increases indicate non-Markovian
information flow \cite{Breuer2009}, in our case from the vibrational degrees
of freedom back to the electronic degrees of freedom. In terms of the slope
of $D$, i.e. $\sigma \left( \rho _{1}\left( t\right) ,\rho _{2}\left(
t\right) \right) =\frac{d}{dt}D\left( \rho _{1}\left( t\right) ,\rho
_{2}\left( t\right) \right) ,$ this implies that $\sigma >0$ for these time
intervals. Formally, a measure for non-Markovianity is defined by the
following \cite{Breuer2009}:%
\begin{equation}
\text{NM-ity}=\max_{\rho _{1}\left( 0\right) ,\rho _{2}\left( 0\right)
}\int_{0}^{\infty }dt~\mathcal{I}\left( \sigma \left( t\right) \right)
~\sigma \left( \rho _{1}\left( t\right) ,\rho _{2}\left( t\right) \right) .
\label{eqNonMarkovianity}
\end{equation}%
Here, the indicator function $\mathcal{I} \left( x>0\right) =1$ $\left( 
\mathcal{I} \left( x<0\right) =0\right)$ monitors only the increasing part
of the trace distance time evolution i.e. the information backflow. The
optimization of the initial states in Eq. (\ref{eqNonMarkovianity}) obtains
the maximally possible non-Markovianity of a particular quantum evolution.
In the case of photosynthetic energy transfer, initial states are given by
the particular physical situation and lead to smaller values for the
non-Markovianity. We use both physical and optimized initial states in this
work.

\textit{Hierarchy equation of motion}$-$ The Hamiltonian describing a single
exciton in a complex with $N$ molecules is given by $H_{e}=\sum_{m=1}^{N}(%
\epsilon _{m}+\lambda )|m\rangle \langle m|+\sum_{m<n}J_{mn}\left( |m\rangle
\langle n|+|n\rangle \langle m|\right) $. The site energies $\epsilon _{m}$,
and couplings $J_{mn}$ are obtained from detailed quantum chemistry studies
and/or fitting of experimental data. The set of states $|m\rangle $ denotes
the site basis. The reorganization energy $\lambda $ is the energy
difference of the non-equilibrium phonon state after Franck-Condon
excitation and the equilibrium phonon state and is assumed to be the same
for each site. The superoperator corresponding to H$_{e}$ is $\mathcal{L}%
_{e}\rho =\left[ H_{e},\rho \right] $. The Hamiltonian for the phonon
environment is $H_{\mathrm{ph}}=\sum_{i}\hbar \omega _{i}\left(
p_{i}^{2}+q_{i}^{2}\right) /2,$ where $\omega _{i,}~p_{i},$ and $q_{i}$ are
the frequency, dimensionless momentum and position operators of the phonon
mode $i.$ The coupling of the electronic degrees of freedom to the phonons
is given by the Hamiltonian $H_{\mathrm{ex-ph}}=\sum_{m}V_{m}\left(
\sum_{i}g_{i}q_{i}\right) _{m}$, with $V_{m}=|m\rangle \langle m|$ and where
we assume that site-energy fluctuations dominate, the bath degrees of
freedom at each site are uncorrelated, and the coupling constants to the
respective modes are given by the constants $g_{i}$. The spectral density,
which describes the coupling strength of exciton to particular phonon modes,
is given by $J\left( \omega \right) =2\lambda \gamma \omega /\left( \omega
^{2}+\gamma ^{2}\right) $. Perhaps the most relevant parameter$,$ the bath
dissipation rate $\gamma $, is related to the bath correlation time by $\tau
_{c}=1/\gamma .$ The equation of motion for the reduced single-exciton
density matrix $\rho $ is obtained by tracing out the less relevant phonon
degrees of freedom and utilizing Wick's theorem for the Gaussian
fluctuations, leading to a non-perturbative hierarchy of system and
auxilliary density operators (ADOs) \cite{Ishizaki2009a,Ishizaki2009}:%
\begin{eqnarray}
\frac{\partial }{\partial t}\sigma ^{\mathbf{n}}\left( t\right)  &=&-\left( i%
\mathcal{L}_{e}+\sum_{m=1}^{N}n_{m}\gamma \right) \sigma ^{\mathbf{n}}\left(
t\right)   \label{eqHierarchyEquation} \\
&&+\sum_{m=1}^{N}\phi _{m}\sigma ^{\mathbf{n}_{m+1}}\left( t\right)
+\sum_{m=1}^{N}n_{m}\theta _{m}\sigma ^{\mathbf{n}_{m-1}}\left( t\right) . 
\notag
\end{eqnarray}%
The system density matrix is the first member of the hierarchy $\rho
(t)=\sigma ^{\mathbf{0}}\left( t\right) .$ The other members of the
hierarchy are arranged in tiers and indexed by $\mathbf{n}=\left(
n_{1,}.....,n_{N}\right) \geq 0$, where a single tier is given by all
elements where $\sum_{m}n_{m}$ is constant. The notation $\mathbf{n}_{m\pm
1}=\left( n_{1,}...,n_{m}\pm 1,...,n_{N}\right) $ is introduced. The
superoperators in the high temperature regime $\hbar \gamma \beta <1$ are $%
\phi _{m}\sigma ^{\mathbf{n}}=i\left[ V_{m},\sigma ^{\mathbf{n}}\right] $
and $\theta _{m}\sigma ^{\mathbf{n}}=i\left( 2\lambda /\beta \hbar ^{2}\left[
V_{m},\sigma ^{\mathbf{n}}\right] -i\lambda \gamma /\hbar \left\{
V_{m},\sigma ^{\mathbf{n}}\right\} \right) $ and act on system and ADOs,
where $\left\{ ,\right\} $ is the anticommutator. Initially, all the
hierarchy members are zero, which corresponds to the electronic ground state
phonon equilibrium configuration. For numerical propagation, the hierarchy
is truncated based on the criterion $\sum_{m}n_{m}\gg \frac{\omega _{e}}{%
\gamma }$, where $\omega _{e}$ is a characteristic frequency of $\mathcal{L}%
_{e}$. We assume that this condition is satisfied when the $\sum_{m}n_{m}$
is by a factor $5$ larger than $\frac{\omega _{e}}{\gamma }$.

The representation of the hierarchy equation of motion in Eq. (\ref%
{eqHierarchyEquation}) is not unique. It can be advantageous to expand the hierarchy
in a set of \textit{normalized} auxiliary systems \cite{Shi2009}. Redefining
the ADOs as $\tilde{\sigma}^{\mathbf{n}}\left( t\right) =\left(
\prod_{m}n_{m}|c_{0}|^{-n_{m}}\right) ^{-1/2}\sigma ^{\mathbf{n}}\left(
t\right) $ and rescaling the superoperators as $\phi _{m}\sigma ^{\mathbf{n}%
_{m+1}}=\sqrt{\left( n_{m}+1\right) |c_{0}|}\phi _{m}\tilde{\sigma}^{\mathbf{%
n}_{m+1}}$ and $n_{m}\theta _{m}\sigma ^{\mathbf{n}_{m-1}}=\sqrt{%
n_{m}/|c_{0}|}\theta _{m}\tilde{\sigma}^{\mathbf{n}_{m-1}}$, with $%
c_{0}=\left( 2\lambda /\beta \hbar ^{2}-i\lambda \gamma /\hbar \right) $ in
the high temperature regime, leads to a more balanced EOM and can lead to a
lower requirement in terms of the number of ADOs. The rescaled
representation is useful for simulating large systems such as the FMO complex 
and for studying the ADOs themselves, as follows.

To characterize the information flow from the point 
of view of the non-Markovian degrees of freedom, 
we propose another measure based on the trace distance. The auxiliary systems
describe perturbations on the time-evolved excitonic state due to the phonon
environment. This separation into system and auxiliary degrees of freedom
implies a quantification of the information content of the auxiliary systems
at a given point in time:%
\begin{equation}
D_{\mathrm{ADO}}^{\mathrm{tot}}\left( t\right) =\sum_{\mathbf{n}\neq \mathbf{%
0}}D\left( \tilde{\sigma}_{1}^{\mathbf{n}}\left( t\right) ,\tilde{\sigma}%
_{2}^{\mathbf{n}}\left( t\right) \right) .  \label{eqADODistance}
\end{equation}%
This measure depends on the electronic initial states and, importantly, uses
the normalized ADOs. Although the $\tilde{\sigma}^{\mathbf{n}}\left(
t\right) $ are not quantum states for $\mathbf{n}\neq \mathbf{0}$, the trace
distance nevertheless measures differences of these operators 
arising from the different initial states. This measure could in principle be
extended to other families of master equations that employ auxiliary degrees of freedom.

\begin{figure}[tbp]
\includegraphics[scale=0.8]{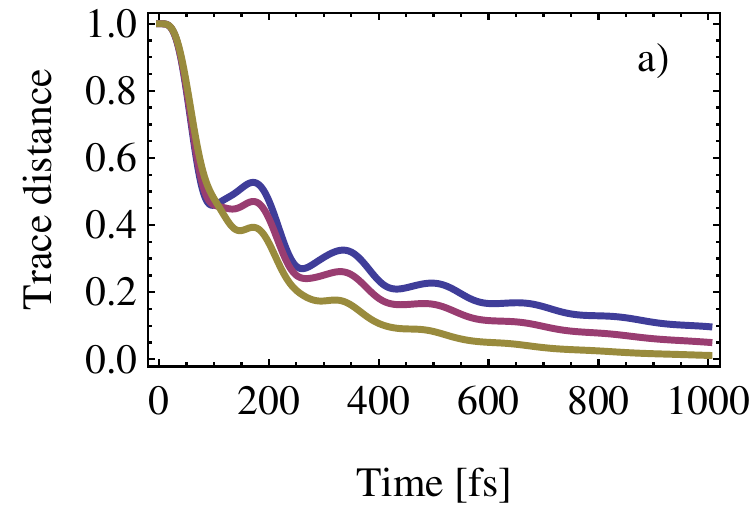} %
\includegraphics[scale=0.8]{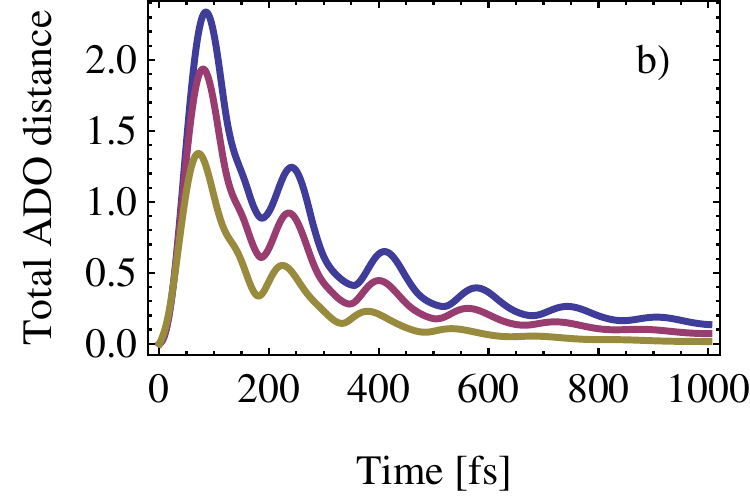} %
\caption{(a) Trace distance of the states $\protect\rho_{1}=|1\rangle
\langle 1|$ and $\protect\rho _{2}=|2\rangle \langle 2|$ as a function of
time for a two-site system (dimer) in the hierarchy equation of motion
model. The trace distance evolves from maximally distinguishable ($=1$) for
the initial states to indistinguishable ($=0$) for the thermal equilibrium
state. At intermediate times, the trace distance increases, which is due to
a reversal of the information flow. Various bath correlation times are
shown, $150$ fs (blue), $100$ fs (red), and $50$ fs (yellow). For long
correlation times, that is, a bath with more memory, the regions of
increasing trace distance are more pronounced. (b) The ADO distance $D_{%
\mathrm{ADO}}^{\mathrm{tot}}$ illustrates the information content in the
non-Markovian degrees of freedom of this model for the same system, initial
states, and bath correlation times as in (a). The anticorrelated
oscillations of the two figures clearly show the information flow between
the system and the NM degrees of freedom. }
\label{figTraceDistanceDimer}
\end{figure}

\textit{Results for a dimer}$-$ In this section, a two-molecule system, or
dimer, is studied in terms of the measure for non-Markovianity, using, if not
otherwise mentioned, the regular hierarchy equation of motion approach Eq. (%
\ref{eqHierarchyEquation}) and scanning over the parameters of the model.
As standard parameters, we use the site energies $\epsilon _{1}=0, $ $%
\epsilon _{2}=120/$cm, and\ the coupling $J=-87.7/$cm$.$ This corresponds to
the strongly-coupled bacteriochlorophyll site 1 and 2 subsystem of the
Fenna-Matthews-Olson complex. Along the lines of the discussion in \cite%
{Ishizaki2009}, the standard bath correlation time is taken from $50$ fs to $%
150$ fs. We assume ambient temperature $T=288~$K$~\left(200/\text{cm}%
\right) .$ The main results use $40$ tiers ($819$ ADOs), and Eq. (\ref%
{eqHierarchyEquation}) is integrated up to maximally $20$ ps.

First, in Fig. \ref{figTraceDistanceDimer} (a), we investigate the time
evolution of the trace distance in Eq. (\ref{eqTraceDistance}) for the dimer
system at reorganization energy $\lambda =20/$cm. We choose the two initial
states $\rho _{1}=|1\rangle \langle 1|$ and $\rho _{2}=|2\rangle \langle 2|$
and the standard parameters given above, while varying the bath dissipation
rate. One can see clear non-Markovian revivals in the time intervals at
around $150$ fs and around $300$ fs, borne out by increases in the trace
distance. For larger correlation time of the bath of $\tau _{c}=150$ fs
these revivals are more pronounced. A bath excitation is more likely to
return to the system instead of being dissipated away rapidly, as occurs in
the case of smaller correlation time $\tau _{c}=50$ fs.

In Fig. \ref{figTraceDistanceDimer} (b), we investigate the time evolution
of the ADO distance measure in Eq. (\ref{eqADODistance}) for the same
system, using the normalized auxiliary systems. At small times, the ADOs are
all zero, thus $D_{\mathrm{ADO}}^{\mathrm{tot}}=0,$ while at long times the
system and ADOs converge to the thermal equilibrium and all information
about the initial states is dissipated. At intermediate times, one finds
strong information flow between system and the non-Markovian degrees of freedom, 
which is borne out by the
anticorrelated oscillations in (a) and (b), respectively. For this, note the
dips at around $150$ fs and around $300$ fs in (b). Additionally, the information
content in the ADOs is smaller when the bath dissipation rate is larger.

The computation of the non-Markovianity measure in Eq. (\ref%
{eqNonMarkovianity}) involves an optimization of the initial states. On the
one hand, the following results use the two initial states $\rho
_{1}=|1\rangle \langle 1|$ and $\rho _{2}=|2\rangle \langle 2|$ as above,
but one the other hand, compare to a systematically-optimized pair of
initial states. One initial state is chosen out of $50$ pure states on the
Bloch sphere. This leads to a total number of $1225$ independent pairs of
states in the optimization. (The state optimization itself is performed with 
$20$ tiers of ADOs.)

\begin{figure}[tbp]
\includegraphics[scale=0.55]{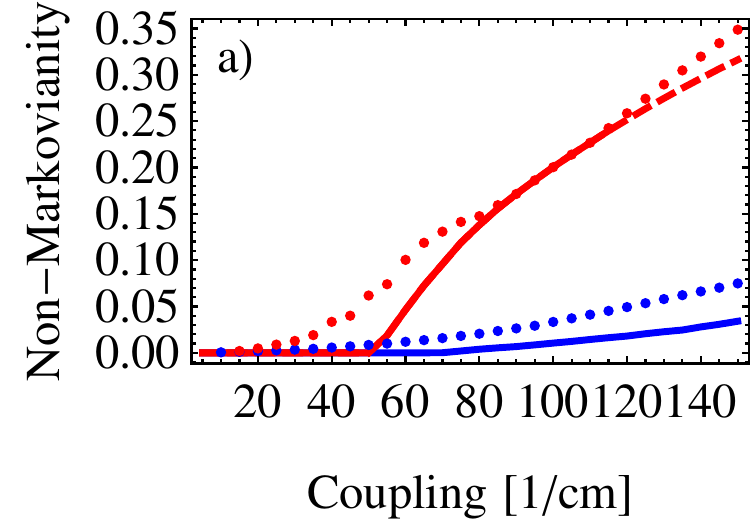} %
\includegraphics[scale=0.55]{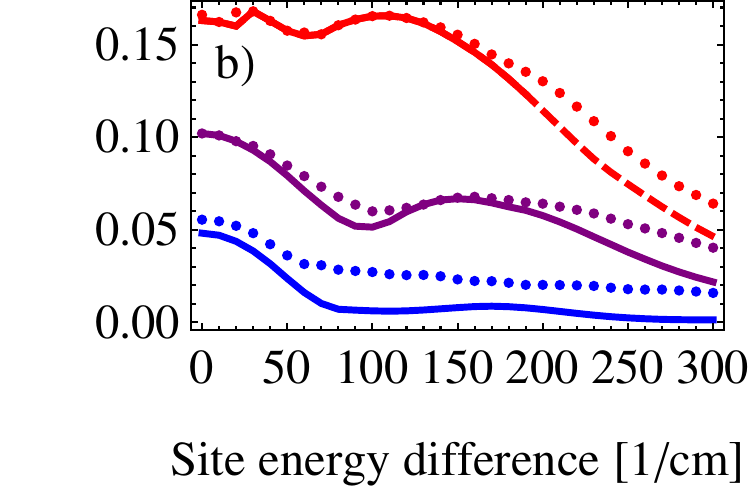} \\%
\includegraphics[scale=0.55]{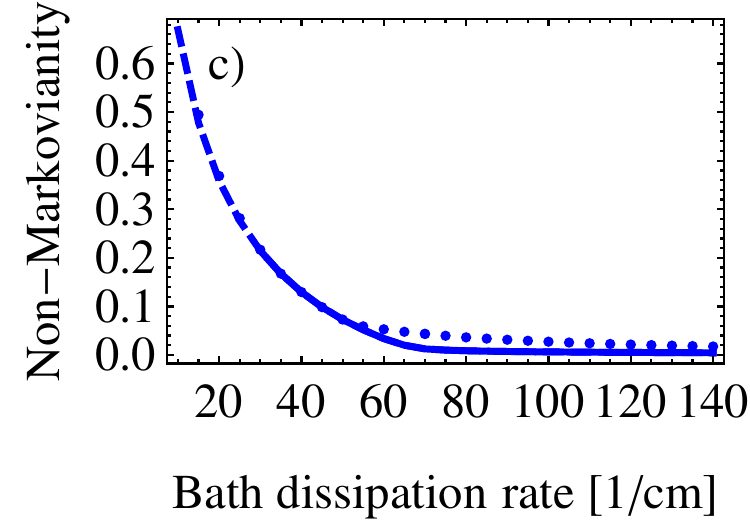} %
\includegraphics[scale=0.55]{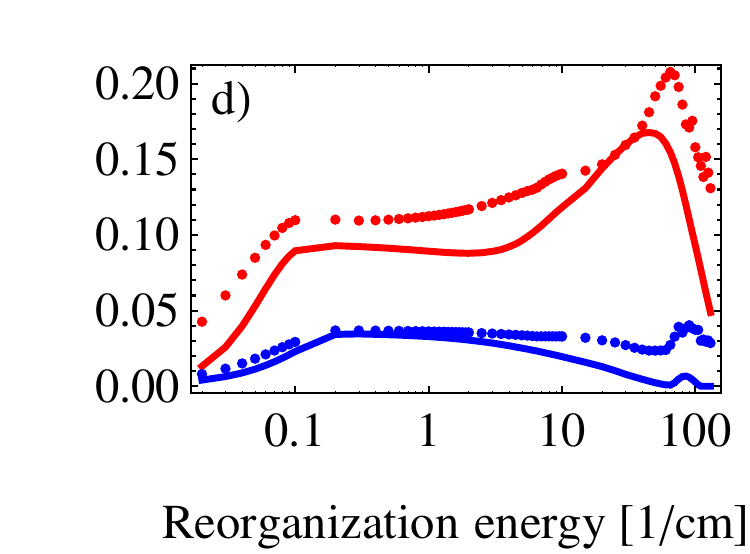} \\%
\includegraphics[scale=0.55]{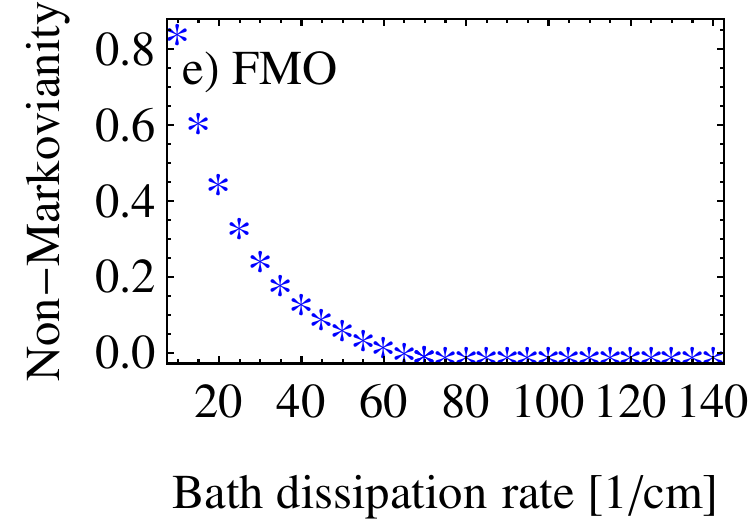} %
\includegraphics[scale=0.55]{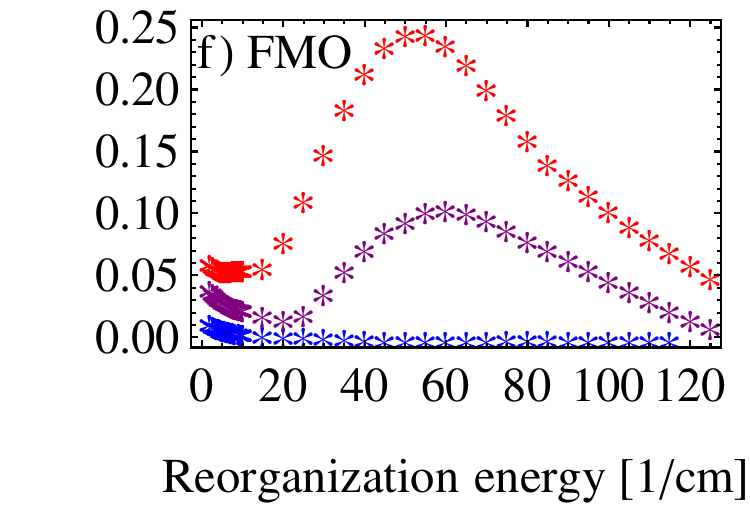}
\caption{ Non-Markovianity as a function of the parameters of a dimer
system, a)-d), and the Fenna-Matthews-Olson complex, e) and f), in the
hierarchy equation of motion approach. For the dimer: a) coupling $J_{12}$,
b) site energy difference $\protect\epsilon_2-\protect\epsilon_1$, c) bath
dissipation rate $\protect\gamma$ (which is related to the bath correlation
time by $\protect\gamma=1/\protect\tau_{c}$,) and d) reorganization energy $%
\protect\lambda$. The solid lines are calculations performed with the
initial states being $\protect\rho_{1}=|1\rangle \langle 1|$ and $\protect%
\rho _{2}=|2\rangle \langle 2|$ while the dots are calculations performed
with optimized initial states. The dashed part of the lines indicates that a
parameter was scanned beyond the validity of the equation based on the
truncation condition. For the FMO complex: e) bath dissipation rate and f)
reorganization energy. In all figures, except in c) and e), we use the
different bath correlation times: $50$ fs (blue), $100$ fs (purple), and $%
150 $ fs (red). }
\label{figNonMarkovianity}
\end{figure}

We now proceed to scan the NM-ity measure 
over the parameters of the model, beginning with the
system parameters, the site energy difference and the chromophoric coupling.
First, the NM-ity is strongly dependent on the chromophoric coupling, see
Fig. \ref{figNonMarkovianity} (a). At zero coupling, the molecules are
independent and the NM-ity is zero; the two initial states remain perfectly
distinguishable. At non-zero coupling, the dynamics shows increasing NM
revivals, while both initial states converge to the same thermal equilibrium
state. Second, with respect to the site energy difference, the dependence of
the NM-ity weakly slopes downward, see Fig \ref{figNonMarkovianity} (b). As
the energy difference increases the coupling becomes less significant and
the molecules become more independent. This leads to decreased NM-ity.

Next, we investigate the role of the bath parameters. First, as mentioned
before, the bath correlation time crucially affects the memory of the bath
and the information back flow to the system. In Fig. \ref{figNonMarkovianity}
(c), we show the NM-ity measure as a function of the inverse bath
correlation time, i.e. the bath dissipation rate. A longer $\tau _{c}$ leads
to more non-Markovianity, because a sluggish bath keeps the memory of
excitations much longer and the information is able to return back to the
electronic degrees of freedom. Second, the other significant bath parameter,
the reorganization energy, is a measure of the non-equilibrium character of
the phonon bath in the excited state. In Fig. \ref{figNonMarkovianity} (d),
we show the NM measure as a function of the reorganization energy. At zero $%
\lambda ,$ which essentially means no coupling to the phonons, the unitary
dynamics obviously shows zero non-Markovianity. At weak $\lambda $, we
observe a NM-ity of around $0.1$ at $t_{c}=150$ fs, which is completely
neglected by Redfield theory. At intermediate $\lambda $ of around $40/$cm,
close to the physiological values of the Fenna-Matthews-Olson complex, the
NM-ity is maximal, showing a value of around $0.2$. At large $\lambda $, the
regime of incoherent transport \cite{Ishizaki2009a}, the NM-ity vanishes.

\textit{Results for the Fenna-Matthews-Olson complex}$-$ We now investigate
the exciton-phonon information flow in the Fenna-Matthews-Olson complex. The
FMO complex is found in green sulphur bacteria, where it connects the
antenna complex with the reaction center during the energy transfer process. 
The FMO complex is a trimer with seven
bacteriochlorophyll (BChl) molecules in each subunit and with one additional
BChl molecule shared between the units. We implement the hierarchy equation
of motion for the seven sites ($N=7$), using the method for rescaling the
auxilliary systems as discussed above and in \cite{Shi2009}. As was shown
recently \cite{Zhu2010}, with four tiers ($329$ ADOs) the coherent
population dynamics can be accurately simulated, but due to the reduced
number of ADOs, we expect to minimally overestimate the NM-ity. We use the
initial states $\rho _{1}=|1\rangle \langle 1|$ and $\rho _{2}=|2\rangle
\langle 2|$. First, the scan of $\gamma $ obtains a strong dependence
similar to the dimer system, see Fig. \ref{figNonMarkovianity} (e). Second,
the scan of the reorganization energy obtains a maximum at around $\lambda
=55/$cm with the NM-ity being $0.2$ at $\lambda =35/$cm and $\tau _{c}=150$
fs, see Fig. \ref{figNonMarkovianity} (f). The results suggest that
non-Markovian effects should play a significant role in the function of the
Fenna-Matthews-Olson complex.

\textit{Conclusion}$-$ The theoretical and experimental characterization of
photosynthetic light-harvesting complexes in terms of non-Markovianity, as
e.g. defined in the recent work \cite%
{Breuer2009,Wolf2008,*Rajagopal2010,*Rivas2010}, can lead to a greater
understanding of the exciton dynamics in these systems. In principle, the
recently proposed ultrafast quantum process tomography protocol extracts a
full characterization of the quantum map \cite{Yuen-Zhou2010}, and could
thus be used for the experimental side of this task. However, experiments
specifically designed to extract a particular observable, such as in our
case the NM-ity measure, could be carried out with a reduced number of experimental
requirements. Future work will investigate analytically and numerically how
to efficiently extract the non-Markovianity from the phase-matched signal
observed in four-wave mixing experiments.

To summarize, we have shown numerically that there is a considerable
non-Markovian information flow between electronic and phononic degrees of
freedom of light-harvesting chromophoric systems under physiological
conditions, utilizing a realistic model for systems such as the
Fenna-Matthews-Olson complex. The results suggest that non-Markovianity
plays a significant role in photosynthetic energy transfer.

The authors are grateful to J. Piilo, C. Rodriguez-Rosario, S. Saikin, and
J. Zhu for insightful discussions. The authors acknowledge funding from DOE,
DARPA, and computational resources through Harvard Research Computing.

\bibliographystyle{apsrev4-1}
\bibliography{PatricksPapers}

\end{document}